# DC and RF Measurements of Serial Bi-SQUID Arrays

G. V. Prokopenko, O. A. Mukhanov, *Fellow, IEEE*, A. Leese de Escobar, *Member, IEEE,* B. Taylor, M. C. de Andrade, *Member, IEEE,* S. Berggren, *Member, IEEE,* P. Longhini, *Member, IEEE,* A. Palacios, M. Nisenoff, *Life Fellow, IEEE,* R. L. Fagaly, *Fellow, IEEE*

*Abstract*—SQUID arrays are promising candidates for low profile antennas and low noise amplifier applications. We present the integrated circuit designs and results of DC and RF measurements of the wideband serial arrays based on integration of linear bi-SQUID cells forming a Superconducting Quantum Interference Filter (bi-SQUID SQIF). Various configurations of serial arrays designs are described. The measured linearity, power gain, and noise temperature are analyzed and compared. The experimental results are matched to results of mathematical modeling. A serial bi-SQUID SQIF arrays are mounted into a coplanar waveguide (CPW) and symmetrically grounded to corresponding sides of CPW. The RF output comes out from the central common line, which is also used for DC biasing and forms a symmetrical balanced output. The signal and DC flux biasing line is designed as coplanar lines passed in parallel over each bi-SQUID cell in a bidirectional fashion concentrating magnetic flux inside of each cell. Serial bi-SQUID SQIF arrays are fabricated on 5 mm x 5 mm chips using standard HYPRES niobium 4.5 kA/cm$^2$ fabrication process.

*Index Terms*—noise temperature, broadband, electrically small antenna, low noise amplifier, high sensitivity, high linearity

## I. INTRODUCTION

SINCE THE INTRODUCTION of Superconducting Quantum Interference Filters (SQIFs) in 2000 [1], [2], the interest in arrays of SQUIDs and their applications has substantially increased. In general, arraying SQUIDs allows one to improve the flux-to-voltage transfer factor, voltage swing, noise characteristics, and dynamic range in contrast to a single SQUID [3]-[5]. Arraying SQUIDs with non-equal loops forms SQIF arrays with a voltage response featuring a single anti-peak at zero magnetic field. This allows new functionalities, such as an ability to measure an absolute magnetic field and improve noise immunity [6], [7]. SQUID arrays and SQIFs, in particular, generated a considerable interest due to the possibility to enhance the core advantages of single SQUIDs with wider bandwidth, higher linearity and dynamic range enabling wider application spectrum. While many applications of SQUID arrays and SQIF arrays were discussed [6]-[18], we focus here on low-noise amplifiers and electrically small antennas.

These have a potential to be used beyond traditional areas of SQUID applications in medical, geomagnetic, and scientific devices for sensitive measurements. Various civilian and defense radio frequency (RF) systems are now being considered for the SQUID arrays to work as a compact front-end device integrated with conventional receivers or as an analog pre-processing module followed by wide-band superconducting Digital-RF receivers being productized today [19]-[25]. In the latter case, the wide bandwidth, low power dissipation, high linearity matches well to characteristics of available superconducting analog-to-digital converters (ADCs) [26]-[31] which otherwise would be constrained by conventional analog electronics.

One of the priority objectives in optimization and synthesis of SQUID arrays is the maximization of its linearity. Many different approaches are pursued in order to improve the linearity of SQIF arrays [32]-[37]. Recently a new approach has emerged based on the linearization of a dc SQUID cell itself. The dc SQUID linearization was accomplished by adding in parallel a non-linear inductance of Josephson junction [38]-[40]. The resultant tri-junction SQUID or bi-SQUID has a distinct triangular flux-voltage response in contrast to a familiar sinewave-shaped response of conventional dc SQUIDs. However, extending the excellent linearity of individual bi-SQUID cells to SQIF arrays was not trivial due to a strong dependence on the values of the additional (3$^{rd}$) junction [38]-[40].

Recently, we have found a method to construct serial bi-SQUID-based SQIF arrays with high linearity [41]. Several arrays were built and their initial measurements showed a good match with our theoretical calculations.

In this paper, we present the results of further experimental studies of serial bi-SQUID SQIF arrays. Specifically, we investigate their linearity, voltage swing, noise properties, gain and bandwidth as function of the number the arrayed bi-SQUID cells. We also present results of experimental investigation of parallel configurations of serial bi-SQUID arrays.

## II. SERIES BI-SQUID ARRAY DESIGN

Fig. 1 shows a schematic of serially connected bi-SQUID cells forming a bi-SQUID SQIF array. Josephson junctions J1

Manuscript received October 9, 2012. This work is supported in part by the Tactical SIGINT Technology Program N66001-08-D-0154, SPAWAR SBIR contracts N00039-08-C-0024, N66001-09-R-0073, ONR (Code 30), the SPAWAR internal research funding (S&T) program and HYPRES IR&D.

G. Prokopenko and O. Mukhanov are with HYPRES, Inc., Elmsford, NY 10523 USA (phone: 914-592-1190; fax: 914-347-2239; e-mail: georgy@hypres.com, mukhanov@hypres.com).
A. Leese de Escobar, P. Longhini, B. Taylor, M. de Andrade are with SPAWAR SSC, San Diego, CA 92152 USA. (e-mail: anna.leese@navy.mil).
S. Berggren and A. Palacios are with SDSU, San Diego, CA 92182 USA (e-mail: susan_berggren@yahoo.com).
M. Nisenoff is with M. Nisenoff Associates, Minneapolis, MN 55403 USA (email: m.nisenoff@ieee.org).
R. L. Fagaly is with Quasar Federal Systems, San Diego, CA 92121 USA (email: fagaly@ieee.org).

and J2 are critically shunted to have $\beta_c \sim 1$; J3 is an additional junction acting as a non-linear inductor. As J3 is not projected to switch during operation, a shunting resistor may not be necessary. However, it is generally advisable to have a resistor shunting dc SQUID for stability [42]. For comparison, we designed serial arrays with and without J3 shunting. As the linearity of bi-SQUID response is a strong function of J3 critical current, we kept J3 unchanged and varied inductors and inductive coupling between control line and bi-SQUID cells guided by theoretical model described in detail earlier [41]. This method is summarized below.

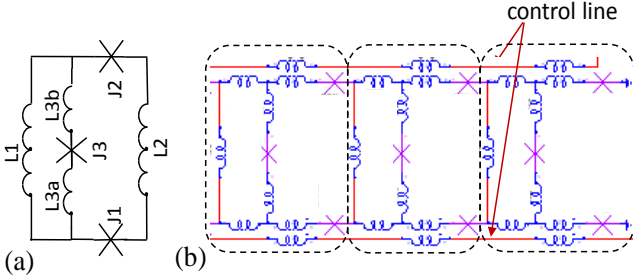

Fig. 1. Schematic of a bi-SQUID SQIF array: (a) bi-SQUID cell; (b) fragment of serial of bi-SQUID array. J1=J2 = J3 = 0.25 mA, $R_{shunt}$ = 2.4 Ohm, $V_c$ = $I_cR_{sh}$ = 600 µV. Circuits with normal (Gaussian) coupling inductance distribution with σ from 25% to 70% were simulated.

### A. Array Modeling

We consider first a single biSQUID as is depicted in Fig. 1(a) and assume the junctions to be identical. Then direct calculations, details of which can be found in [41], lead to the following system of Ordinary Differential Equations that together govern the phase dynamics across each junction:

$$(L_1 + \tfrac{1}{2}L_2)\dot{\phi}_1 - \tfrac{1}{2}L_2\dot{\phi}_2 - L_1\dot{\phi}_3 = \tfrac{1}{2}L_1 i_b + \phi_2 - \phi_1 + 2\pi\phi_e a_n + L_1 i_{c3}\sin\phi_3 \\ + \tfrac{1}{2}L_2\sin\phi_2 - (L_1 + \tfrac{1}{2}L_2)\sin\phi_1 \quad (1)$$

$$\tfrac{1}{2}L_2\dot{\phi}_1 - (L_1 + \tfrac{1}{2}L_2)\dot{\phi}_2 - L_1\dot{\phi}_3 = -\tfrac{1}{2}L_1 i_b + \phi_2 - \phi_1 + 2\pi\phi_e a_n + L_1 i_{c3}\sin\phi_3 \\ - \tfrac{1}{2}L_2\sin\phi_1 + (L_1 + \tfrac{1}{2}L_2)\sin\phi_2$$

$$\tfrac{1}{2}L_2\dot{\phi}_1 - \tfrac{1}{2}L_2\dot{\phi}_2 - (L_{3a} + L_{3b})\dot{\phi}_3 = \phi_2 - \phi_3 + \phi_1 - (L_{3a} + L_{3b})i_{c3}\sin\phi_3 \\ - \tfrac{1}{2}L_2\sin\phi_1 + \tfrac{1}{2}L_2\sin\phi_2,$$

where $\phi_i$ are the phases on each of the junctions $J_n$, $n = 1,2,3$, $i_{c3} = I_{c3}/I_c$, is the normalized critical current across the third junction $J_3$, $I_{c1} = I_{c2} = I_c$, $a_n$ is a nonlinearity parameter related to the loop size between $J_1$ and $J_2$, and ( )denotes differentiation with respect to the normalized time $\tau = \omega_c t$, $\omega_c = 2\pi V_c/\Phi_0$, $V_c = I_c R_N$.

Equations (1) form the building block for modeling the phase dynamics across each junction of a serially connected array of $N$ biSQUIDs, as is sketched in Fig. 1(b). The derivation of the equations governing the phase across each of the junctions is technically tedious. The end result yields a system of equations that mimics the governing equation (1) of each individual biSQUID with a nearest neighbor coupling term. For brevity we list in here the system of equations and refer the readers to [41] for more details of the derivation.

$$(L_{1,i} + \tfrac{1}{2}L_{2,i})\dot{\phi}_{i,1} - \tfrac{1}{2}L_{2,i}\dot{\phi}_{i,2} - L_{1,i}\dot{\phi}_{i,3} = \tfrac{1}{2}L_{1,i} i_b + \phi_{i,2} - \phi_{i,1} + 2\pi\phi_e a_{n,i} + M\Phi_i \\ + L_{1,i} i_{c3,i}\sin\phi_{i,3} + \tfrac{1}{2}L_{2,i}\sin\phi_{i,2} \quad (2) \\ - (L_{1,i} + \tfrac{1}{2}L_{2,i})\sin\phi_{i,1}$$

$$\tfrac{1}{2}L_{2,i}\dot{\phi}_{i,1} - (L_{1,i} + \tfrac{1}{2}L_{2,i})\dot{\phi}_{i,2} - L_{1,i}\dot{\phi}_{i,3} = -\tfrac{1}{2}L_{1,i} i_b + \phi_{i,2} - \phi_{i,1} + 2\pi\phi_e a_{n,i} + M\Phi_i \\ + L_{1,i} i_{c3,i}\sin\phi_{i,3} - \tfrac{1}{2}L_{2,i}\sin\phi_{i,1} \\ + (L_{1,i} + \tfrac{1}{2}L_{2,i})\sin\phi_{i,2}$$

$$\tfrac{1}{2}L_{2,i}\dot{\phi}_{i,1} - \tfrac{1}{2}L_{2,i}\dot{\phi}_{i,2} - (L_{3a,i} + L_{3b,i})\dot{\phi}_{i,3} = \phi_{i,2} - \phi_{i,3} + \phi_{i,1} + M\Phi_i \\ - (L_{3a,i} + L_{3b,i})i_{c3,i}\sin\phi_{i,3} \\ - \tfrac{1}{2}L_{2,i}\sin\phi_{i,1} + \tfrac{1}{2}L_{2,i}\sin\phi_{i,2},$$

where $\phi_{i,j}$ are the phases on each of the junctions $J_{i,j}$, $i = 1..N$, $j = 1,2,3$, $a_{n,i}$ is a parameter related to the loop size between $J_{i,1}$ and $J_{i,2}$, $i_{c3,i} = I_{c3,i}/I_c$ is the normalized critical current of the third junction $J_{3,i}$ of the $i^{th}$ biSQUID, and $M$ is the coupling strength for the phase interaction $\Phi_i$ between nearest neighbors — one neighbor for the edge elements, two for the inner elements — according to

$$\Phi_i = \begin{cases} \dfrac{1}{a_{n,2}}(\phi_{2,1} - \phi_{2,2} - 2\pi\phi_e a_{n,2}), & \text{for } i = 1 \\ \dfrac{1}{a_{n,i+1}}(\phi_{i+1,1} - \phi_{i+1,2} - 2\pi\phi_e a_{n,i+1}) + & \\ \dfrac{1}{a_{n,i-1}}(\phi_{i-1,1} - \phi_{i-1,2} - 2\pi\phi_e a_{n,i-1}), & \text{for } i = 2,\ldots, N-1 \\ \dfrac{1}{a_{n,N-1}}(\phi_{N-1,1} - \phi_{N-1,2} - 2\pi\phi_e a_{n,N-1}), & \text{for } i = N \end{cases} \quad (3)$$

For simplicity, we assume all inductances to be identical throughout the array. However, the computer code that was written to simulate the voltage response of the array can easily handle the case of non-identical inductances. Following the derivation of the array equations (2), we conducted in [39] extensive numerical simulations to investigate the voltage response of the serial array as a function of the external field $\phi_e$ and the critical current $i_{c3}$. Different distributions of loop sizes were considered for each array, including: linear, Gaussian, exponential, logarithmic, and equal size. Among these distributions, the Gaussian distribution in a serial array produced the highest linear response around the anti-peak. Note that other distributions excluding the equal sized were very similar to the Gaussian, however, the Gaussian was only slightly better and it would be redundant to display results on the other distributions. As expected, the voltage output formed an anti-peak at $\phi_e = 0$ magnetic flux and, more importantly, we were able to manipulate or control the linearity of the anti-peak by changing $i_{c3}$. In fact, for small magnitudes of that parameter the voltage response curve appeared to mimic that of a conventional SQIF device made up of two-junctions per loop. But as $i_{c3}$ increased the linearity of the anti-peak increased as well and, eventually, the voltage response, near zero flux, resembles the triangular shape of the voltage output of a single bi-SQUID. We wish to emphasize that linearity was measured through the least squares error of fitting a line around the anti-peak. As expected, the error in the linear fitting was found to decrease while $i_{c3}$ increased, thus indicating an improvement in the linearity of the voltage output. More importantly, the linear fitting test showed an optimal value of the critical current where Spur Free Dynamic Range is optimum and beyond which only marginal improvements in linearity can be achieved.

*B. Bi-SQUID Layout Design*

Fig. 2 shows examples of layout designs of serial bi-SQUID SQIF arrays produced for fabrication using HYPRES Niobium superconductor integrated circuit fabrication process [43]-[45] with a 4.5 kA/cm$^2$ Josephson junction critical current density. Serial bi-SQUID SQIF arrays are mounted into a coplanar waveguide (CPW) and symmetrically grounded to corresponding sides of CPW. The RF output comes out from the central common line, which is also used for DC biasing and forms a symmetrical balanced output. The signal and DC flux biasing control line is designed as coplanar lines passed in parallel over each bi-SQUID cell in an opposite direction concentrating magnetic flux inside of each cell. We have designed arrays with a different number of flux bias line turns to vary coupling strength.

By comparing Fig. 1(a) and Fig. 2(a), it is evident that inductances L3a and L3b cannot be done equal due to the layout constraints. In order to avoid an accumulation of an inductive skew on each side of the array, we flip every other cell to equalize average inductances across both array sides. All arrays are designed with normal Gaussian distribution of cell inductances of the order of ~70% as guided by numerical modeling described in the previous section. This is achieved by changing each bi-SQUID cell inductance, i.e., a size of cell area as shown in Fig. 2(b).

Figs. 3, 4 show images of the fabricated serial bi-SQUID SQIF arrays. We designed and fabricated a large variety of bi-SQUID SQIF arrays targeting both the low-noise amplifier and antenna applications. In this paper, we present only experimental results obtained in testing arrays as a low-noise amplifier. Fig. 3(a) shows a 5 x 5 mm$^2$ chip containing three serial bi-SQUID SQIF arrays with 20-, 200-, 1000-cell designs in order to experimentally investigate dependences of array characteristics on the number of integrated bi-SQUID cells. The design of the coupled control line (flux bias) is done using a 3-turn line to maximize gain, as the same line is used to input RF signal. The array response is measured at the dc current bias point of the array.

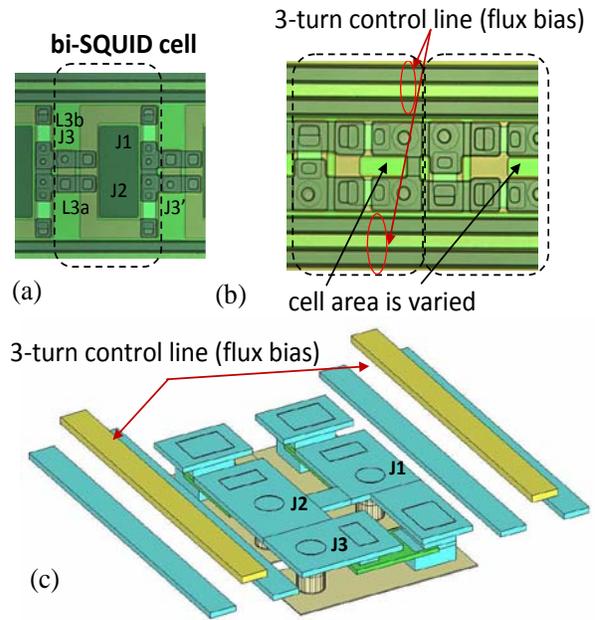

Fig. 2. Layout design of a bi-SQUID SQIF array: (a) bi-SQUID cell used in arrays of Fig. 4; (b) fragment of serial bi-SQUID array used in array of Fig. 3; (c) a 3D sketch of the bi-SQUID cell showing the layer design. Note, adjacent cells are flipped in respect of J3 placement to avoid effect different L3a and L3b accumulation.

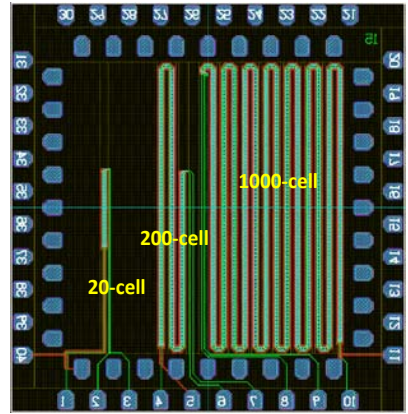

Fig. 3. Chip layout of different three bi-SQUID SQIF arrays with 20, 200, and 1000 cells. A 3-turn flux bias line is used.

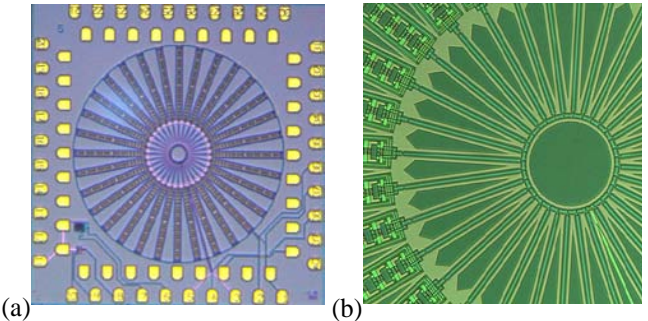

Fig. 4. Microphotographs of fabricated chips of parallel connection of 32 serial bi-SQUID SQIF consisting of 33 cells: (a) 5 x5 mm$^2$ chip with a 32-spoke wheel design; (b) zoom-in of the array central part.

Fig. 4 shows a parallel connection of serial bi-SQUID SQIF arrays arranged in a multi-spoke wheel fashion targeting antenna applications. Several versions with different number of spokes (4, 8, 16, 32) were implemented. The objective of these designs is to investigate shape dependences and effects of mutual coupling between different serial arrays.

### III. EXPERIMENTAL EVALUATION

#### A. DC Measurements

We performed all experimental measurements in liquid helium dewar using HYPRES standard cryoprobe with 40 coaxial lines. For the DC measurements, we constructed a set of experimental setups to measure current-voltage characteristics (IVC), voltage-flux (voltage-flux bias) response, and noise properties. For brevity, we present here only a noise measurement setup (Fig. 5).

Fig. 6(a) shows the comparison of the measured IVC measured for the 20, 200, and 1000 bi-SQUID serial arrays from the chip shown in Fig. 3. Fig. 6(b) shows and example of the 1000-cell array IVC modulation with different values of the flux bias showing significant modulation achieved.

In order to compensate for growing impedance of measured serial arrays with a number of arrayed bi-SQUID cells, we use a dc bias source with switchable output impedances. This is necessary to prevent the loading effect, limiting the measured voltage response as illustrated in Fig. 6(c),(d).

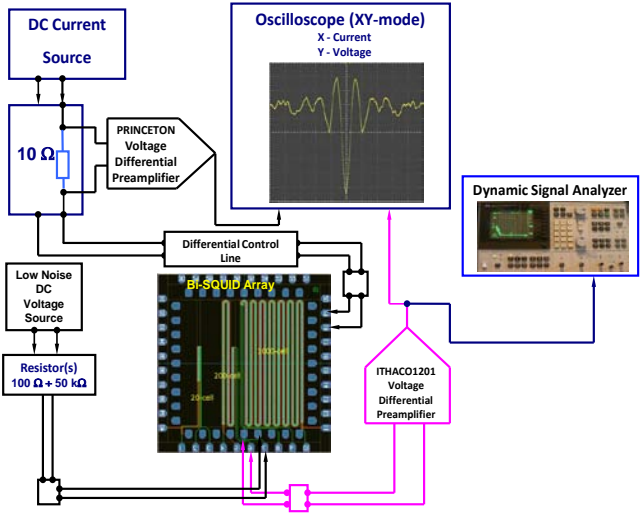

Fig. 5. Example of experimental setup used for noise measurements.

The setup shown in Fig. 5 was used to evaluate the noise characteristics of bi-SQUID SQIF arrays of various lengths. Fig. 7(a) shows the comparison of the measured flux noise spectral densities for the 20- and 1000-cell arrays. It is evident that noise is getting reduced with as $\sim N^{1/2}$ or $\sim 7$ times as was theoretically expected.

From these measurements, we can infer the noise energy and noise temperature follow [46], [47]. The noise energy per unit bandwidth via flux noise in SQUID is defined as:

$$\varepsilon(f) = S_\Phi(f)/2L, \quad (4)$$

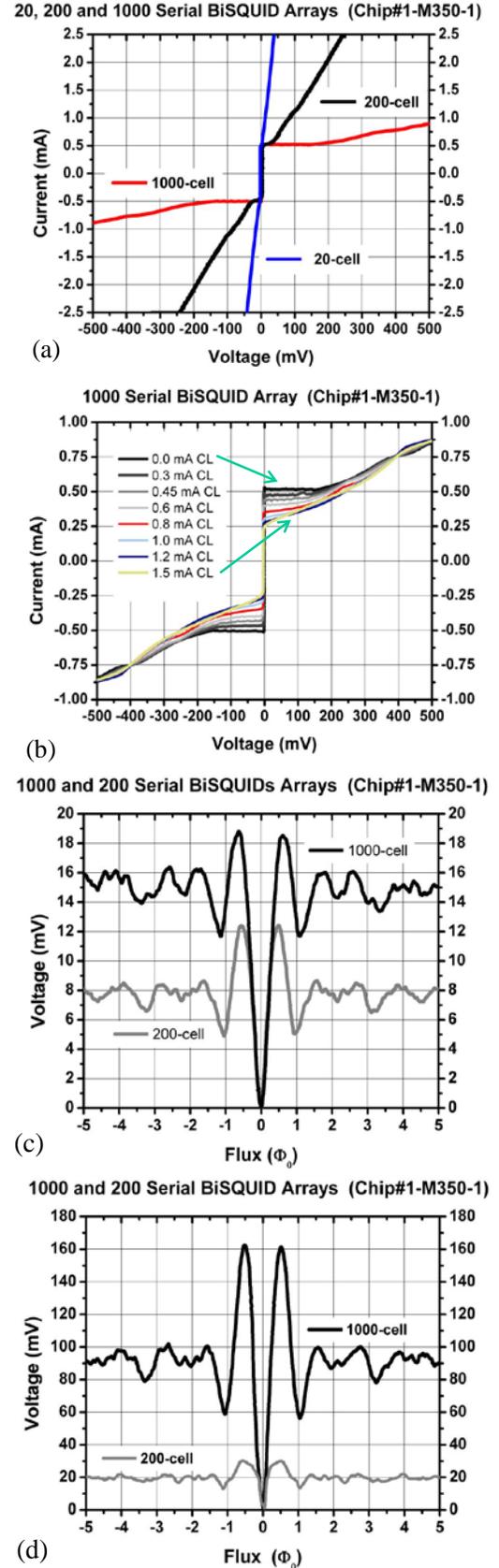

Fig. 6. Measured (a) comparison of measured IVCs for 20, 200, 100-cell arrays; (b) 1000-cell array Ic modulation with flux bias applied via control line (CL). The measured flux/voltage response with (c) 200 Ohm and (d) 5.4 kOhm output current source impedances demonstrating the source loading effect.

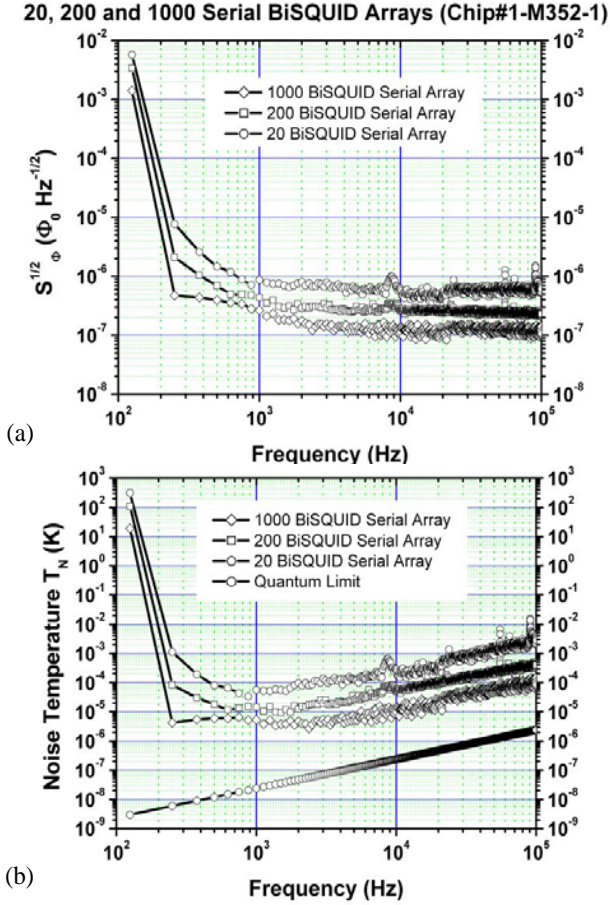

Fig. 7. Measured flux noise spectral density and inferred from this noise temperature for serial 20 and 1000 bi-SQUID SQIF arrays.

where $f$ is frequency, $L$ – inductance of bi-SQUID calculated from the measured separately, $\Delta I_c$ modulation of IV curve defined as $L = \Phi_0/2\Delta I_c$. Noise temperature is defined as

$$T_N = \pi f\, \varepsilon(f)/k_B, \qquad (5)$$

where $k_B$ is Boltzmann's constant. The resultant noise temperature is plotted in Fig. 7(b) for the 20- and 1000-cell arrays.

Similar to serial arrays from chip of Fig. 3, we evaluated a parallel-serial "multi-spoke wheel" configuration depicted in Fig. 4. Fig. 8 shows the measured noise properties of the design. As it is evident from the measured data, its noise properties are worse. We attribute this to a complicated flux bias pattern, as this chip was designed for antenna measurements with magnetic field flux biasing rather the current biasing used in these measurements.

### B. RF Measurements

We performed RF testing in order to evaluate our bi-SQUID SQIF arrays as low-noise amplifiers. Fig. 9 shows the measured output power and power gain over a 50 MHz frequency band (130 to 180 MHz) for serial arrays measured at three array bias points for comparison – at the tip of anti-peak (a reference point with no amplification), at mid-point of the anti-peak slope (a maximum gain point in the linear region) and at the point close to a saturation outside the anti-peak (low gain, non-linear region). At point 1 (close to the array critical current), we applied a small RF power as a reference level via the flux bias line using a sweep RF generator. Then, we increase the array flux bias to move the operation point to the approximately mid-point of the anti-peak slope (with the same input RF power level) observing the output power corresponding to the top trace (trace 2) in Fig. 9(a),(c). Then, we increase the array dc bias further to move the operation point to outside of the anti-peak (trace 3 in Fig. 9.) keeping the same RF power.

The output power is measured directly by a spectrum analyzer without using a pre-amplifier. Using output power at point 1 as a reference, we calculate the power gain at the mid-point and outside of the anti-peak (Fig. 9(b),(d)). The maximum gain was recorded as 22 dB at 168 MHz and 30 dB at 130 MHz for the 200- and 1000-cell arrays, respectively. As it is evident from Fig. 9, the array gain is not flat. We attribute this to the non-optimized, unmatched flux bias line design used to input the RF power.

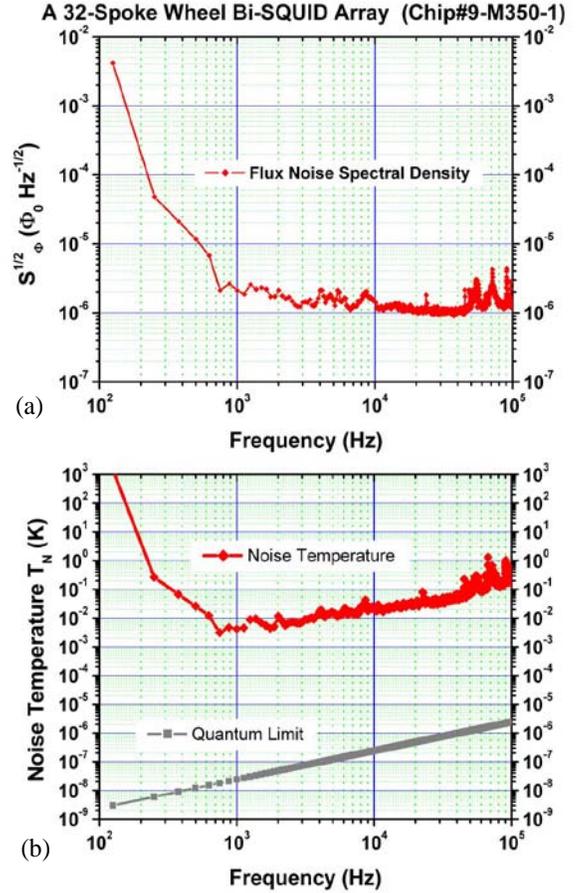

Fig. 8. Measured flux noise spectral density and inferred from this noise temperature for 32-spoke wheel (parallel combination of 32 serial bi-SQUID SQIF arrays) from chip depicted in Fig. 4. Frequency spikes above 20 kHz are attributed to an external noise.

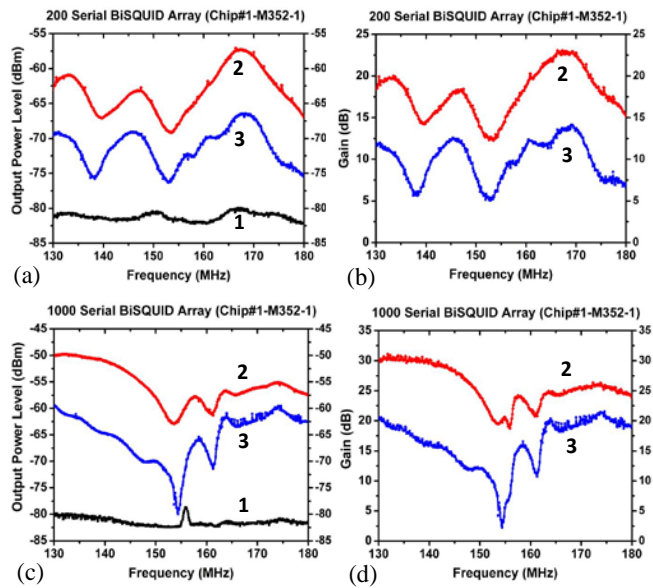

Fig. 9. Measured (a), (c) output power level and (b), (d) gain for a 200- and a 1000-cell serial bi-SQUID arrays at three dc bias levels placing the array operation point at 1 – near the tip of the anti-peak, 2 – a midpoint of the anti-peak slope (optimum), 3 – outside of the anti-peak (saturation region).

In order to evaluate the linearity dependence on the number of array cells, we perform the initial comparative two-tone measurements for 200- and 1000-cell arrays with operation point chosen at the middle of the anti-peak slope. Fig. 10 shows the results of these measurements for 158 and 162 MHz with linearity growing from ~27 dB for the 200-cell array to ~40 dB for the 1000-cell array. For a full analysis, such measurements should be performed for multiple frequencies across the bandwidth.

## IV. Conclusion

We have experimentally studied the DC and RF characteristics of serial SQIF arrays comprised of bi-SQUID cells. The current-voltage characteristics, flux-voltage response, noise properties, linearity, power gain are analyzed and compared.

The test results of serial bi-SQUID SQIF arrays validated theoretical expectations that bi-SQUID arrays should act in a similar to dc SQUID array fashion. We measured the RF properties of serial bi-SQUID arrays including gain, gain flatness, and linearity in a two-tone measurement. From comparison of different size arrays, it is evident that longer arrays produce better results in all aspects of an amplifier performance.

The presented here results should be considered as low-end estimates, as the accuracy of measurements can be further improved with using more sophisticated setup, e. g, with using cold pre-amplifiers and attenuators to block thermal noise from room temperature equipment.

The approach presented here, which is based on using serial bi-SQUID SQIF arrays for low noise amplifiers, can be extended into two-dimensional (2D) bi-SQUID SQIF arrays [48].

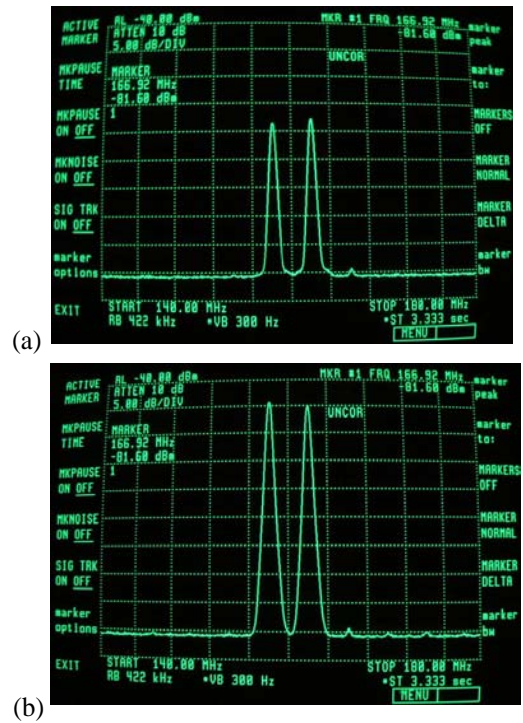

Fig. 10. Measured two-tone (158 MHz and 162 MHz) response for (a) 200-cell and (b) 1000-cell serial bi-SQUID array at dc bias level placing operation point at a midpoint of anti-peak SQIF slope with 5 dB/div.

## Acknowledgment

Authors are grateful to D. Kirichenko, V. Kornev, D. Bowling, J. Talvacchio, J. Przybysz for useful discussions, S. Cybart, A. Matlashov, M. Mueck for advice in testing, V. Dotsenko for help with cryoprobes. Special thanks go to the HYPRES fabrication team of D. Yohannes, J. Vivalda, R. Hunt, D. Donnelly for manufacturing the integrated circuits.

## References

[1] J. Oppenländer, C. Häussler, and N. Schopohl, "Non-Phi(0)-periodic macroscopic quantum interference in one-dimensional parallel Josephson junction arrays with unconventional grating structure", *Phys. Rev. B*, vol. 63, pp. 024511-20, 2000.
[2] C. Häussler, J. Oppenländer, and N. Schopohl "Nonperiodic flux to voltage conversion of series arrays of dc superconducting quantum interference devices," *J. Appl. Phys.* vol. 89 , pp. 1875-1879, 2001.
[3] R. P. Welty and J. M. Martinis, "A series array of DC SQUIDs," *IEEE Trans. Magn.*, vol. 27, pp. 2924-2926, Mar. 1991.
[4] K. G. Stawiasz and M. B. Ketchen, "Noise measurements of series SQUID arrays," *IEEE Trans. Appl. Supercond.*, vol. 3, pp. 1808-1811, Mar. 1993.
[5] D. Drung, "High-Tc and low-Tc dc SQUID electronics," *Supercond. Sci. Technol.*, vol. 16, p. 1320, 2003.
[6] J. Oppenländer, C. Häussler, T. Träuble, and N. Schopohl, "Highly sensitive magnetometers for absolute magnetic field measurements based on quantum interference filters," *Physica C*, vol. 368, pp. 119-124, 2002.
[7] V. Schultze, R. IJsselsteijn, H.-G. Meyer, J. Oppenländer, Ch. Häussler, and N. Schopohl, "High-Tc superconducting quantum interference filters for magnetometers," *IEEE Trans. Appl. Supercond.*, vol. 13, pp. 775-778, 2003.
[8] R. Fagaly, "Superconducting quantum interference device instruments and applications," *Rev. Sci. Instrum.*, vol. 77, 101101, Oct. 2006.
[9] K. Kazami, J. Kawai, G. Uehara, and H. Kado, "A 35-series superconducting quantum interference device array for high-dynamic-

range magnetic measurements," *Jpn. J. Appl. Phys.*, vol. 35, pp. 4322-4326, 1996.

[10] V. K. Kornev, I. I. Soloviev, O. A. Mukhanov, "Possible approach to the driver design based on series SQIF," *IEEE Trans. Appl. Supercond.*, vol. 15, no. 2, pp. 388- 391, Jun. 2005.

[11] V. K. Kornev, I. I. Soloviev, N.V. Klenov, O. A. Mukhanov, "Development of SQIF-based output broad band amplifier," *IEEE Trans. Appl. Supercond.*, vol. 17, no. 2, pp. 569-572, Jun. 2007.

[12] O. V. Snigirev, M. L. Chukharkin, A. S. Kalabukhov, M. A. Tarasov, et al., "Superconducting quantum interference filters as RF amplifiers," *IEEE Trans. Appl. Supercond.*, vol. 17, pp. 718-721, Jun. 2007.

[13] P. Caputo, J. Tomes, J. Oppenländer, Ch. Häussler, et al., "Two-tone response of radiofrequency signals using the voltage output of a Superconducting Quantum Interference Filter," *J. Supercond. and Novel Magnetism*, vol. 20, pp. 25-30, 2007.

[14] V. K. Kornev, I. I. Soloviev, N. V. Klenov, T.V. Filippov, et al., "Performance advantages and design issues of SQIFs for microwave applications," *IEEE Trans. Appl. Supercond.*, vol. 19, pp. 916-919, Jun. 2009.

[15] J. Luine, L. Abelson, D. Brundrett, J. Burch, et al., "Application of a DC SQUID array amplifier to an electrically small active antenna," *IEEE Trans. Appl. Supercond.*, vol. 9, pp. 4141-4144, 1999.

[16] V. K. Kornev, I. I. Soloviev, N. V. Klenov, A. V. Sharafiev, and O. A. Mukhanov, "Linear Bi-SQUID arrays for electrically small antennas," *IEEE Trans. Appl. Supercond.*, vol. 21, no. 3, pp. 713-716, Jun. 2011.

[17] V. Kornev, I. I. Soloviev, N. V. Klenov, A. V. Sharafiev, and O. A. Mukhanov, "Array designs for active electrically small superconductive antennas," *Physica C*, vol. 479, pp. 119–122, Sep. 2012.

[18] V. K. Kornev, I. I. Soloviev, A. V. Sharafiev, and O. A. Mukhanov, "Active electrically small antenna based on superconducting quantum array," *IEEE Trans. Appl. Supercond.*, vol. 23, 2013 (in press); arXiv:1211.6787.

[19] D. K. Brock, O. A. Mukhanov, and J. Rosa, "Superconductor Digital RF development for software radio," *IEEE Communications Mag.*, vol. 39, pp. 174-179, Feb. 2001.

[20] D. Gupta, T. V. Filippov, A. F. Kirichenko, D. E. Kirichenko, et al., "Digital channelizing radio frequency receiver," *IEEE Trans. Appl. Supercond.*, vol. 17, pp. 430-437, Jun. 2007.

[21] O. A. Mukhanov, D. Kirichenko, I. V. Vernik, T. V. Filippov, et al., "Superconductor Digital-RF receiver systems," *IEICE Trans. Electron.*, vol. E91-C, no. 3, pp. 306-317, Mar. 2008.

[22] I. V. Vernik, D. E. Kirichenko, V. V. Dotsenko, R. J. Webber, et al., "Progress in the development of cryocooled digital channelizing RF receivers," *IEEE Trans. Appl. Supercon.*, vol. 19, pp. 1016-1021, Jun. 2009.

[23] A. Leese de Escobar, R. Hitt, O. Mukhanov, W. Littlefield, "High performance HF-UHF all digital RF receiver tested at 20 GHz clock frequencies," in: *Mil. Comm. Conf. (MILCOM'06)*, Washington, DC, USA, Oct. 2006.

[24] J. Wong, R. Dunnegan, D. Gupta, D. Kirichenko, V. Dotsenko, R. Webber, R. Miller, O. Mukhanov, R. Hitt, "High performance, All Digital RF receiver tested at 7.5 GigaHertz," in: *Mil. Comm. Conf. (MILCOM'07)*, Orlando, FL, USA, Oct. 2007.

[25] I. V. Vernik, D. E. Kirichenko, V. V. Dotsenko, R. Miller, R. J. Webber, P. Shevchenko, A. Talalaevskii, D. Gupta, and O. A. Mukhanov, "Cryocooled wideband digital channelizing RF receiver based on low-pass ADC," *Superconductor Science and Technology*, vol. 20, pp. S323-S327, Nov. 2007.

[26] O. Mukhanov, D. Gupta, A. Kadin, and V. Semenov, "Superconductor Analog-to-Digital converters," *Proc. of the IEEE*, vol. 92, pp. 1564-1584, Oct. 2004.

[27] I. V. Vernik, D. E. Kirichenko, T. V. Filippov, A. Talalaevskii, A. Sahu, A. Inamdar, A. F. Kirichenko, D. Gupta, and O. A. Mukhanov, "Superconducting high-resolution low-pass analog-to-digital converters," *IEEE Trans. Appl. Supercond.*, vol. 17, pp. 442-445, Jun. 2007.

[28] A. Sekiya, K. Okada, Y. Nishido, A. Fujimaki, and H. Hayakawa, "Demonstration of the multi-bit sigma-delta A/D converter with the decimation filter," *IEEE Trans. Appl. Supercond.*, vol. 15, pp. 340-343, Jun. 2005.

[29] Q. Herr, D. Miller, A. Pesetski, and J. Przybysz, "A quantum-accurate two-loop data converter," *IEEE Trans. Appl. Supercond.*, vol. 19, pp. 676-679, Jun. 2009.

[30] A. Inamdar, S. Rylov, A. Talalaevskii, A. Sahu, S. Sarwana, et al., "Progress in design of improved high dynamic range analog-to-digital converters," *IEEE Trans. Applied Superconductivity*, vol. 19, pp. 670-675, Jun. 2009.

[31] D. Gupta, A. Inamdar, D. E. Kirichenko, A. M. Kadin, and O. A. Mukhanov, "Superconductor analog-to-digital converters and their applications," *Microwave Symposium Digest (MTT), 2011 IEEE MTT-S*, pp. 1-4, June 2011.

[32] V. K. Kornev, I. I. Soloviev, N. V. Klenov, and O. A. Mukhanov, "Synthesis of high-linearity array structures," *Superconductor Science and Technology*, vol. 20, pp. S362-S366, Nov. 2007.

[33] V. K. Kornev, I. I. Soloviev, N. V. Klenov, and O. A. Mukhanov, "High linearity Josephson-junction array structures," *Physica C*, vol. 468, issue 7-10, pp. 813-816, Apr. 2008.

[34] P. Longhini, A. Leese de Escobar, F. Escobar, V. In, A. Bulsara, "Novel coupling scheme for the dynamics of non-uniform coupled SQUID," *IEEE Trans. Appl. Supercond.*, vol. 19, pp. 749-752, Jun. 2009.

[35] V. K. Kornev, I. I. Soloviev, N. V. Klenov, and O. A. Mukhanov, "High-linearity SQIF-like Josephson-junction structures," *IEEE Trans. Appl. Supercond.*, vol. 19, pp. 741-744, Jun. 2009.

[36] S. A. Cybart, S. M. Wu, S. M. Anton, I. Siddiqi, et al., "Series array of incommensurate superconducting quantum interference devices from YBa2Cu3O7-δ ion damage Josephson junctions," *Appl. Phys. Lett.*, vol. 93, 182502, 2008.

[37] V. K. Kornev, I. I. Soloviev, N. V. Klenov, and O. A. Mukhanov, "Design and experimental evaluation of SQIF arrays with linear voltage response," *IEEE Trans. Appl. Supercond.*, vol. 21, no. 3, pp. 394-398, Jun. 2011.

[38] V. K. Kornev, I. I. Soloviev, N. V. Klenov, and O. A. Mukhanov, "Bi-SQUID: a novel linearization method for dc SQUID voltage response," *Superconductor Science and Technology*, vol. 22, 114011, Nov. 2009.

[39] V. K. Kornev, I. I. Soloviev, N. V. Klenov, and O. A. Mukhanov, "Progress in high-linearity multi-element Josephson structures," *Physica C*, vol. 470, issue 19, pp. 886-889, Oct. 2010.

[40] I. I. Soloviev, V. K. Kornev, N. V. Klenov and O. A. Mukhanov, "Superconducting Josephson structures with high linearity of transformation of magnetic signal into voltage," *Physics of the Solid State*, vol. 52, no. 11, pp. 2252–2258, 2010.

[41] P. Longhini, S. Berggren, A. Leese de Escobar, A. Palacios, S. Rice, et al., "Voltage response of non-uniform arrays of bi-superconductive quantum interference devices," *J. Appl. Phys.*, vol. 111, 093920, May 2012.

[42] J. Knuutila, A. Ahonen, C. Tesche, "Effects on DC SQUID characteristics of damping of input coil resonances," *J. Low Temp. Phys.*, vol. 68, pp. 269-284, 1987.

[43] HYPRES Nb Process Design Rules. Revision #24 Jan 11, 2008. http://www.hypres.com.

[44] D. K. Brock, A. M. Kadin, A. F. Kirichenko, O. A. Mukhanov, et al., "Retargeting RSFQ cells to a submicron fabrication process," *IEEE Trans. Appl. Supercond.*, vol. 11, no. 1, pp. 369-372, Mar. 2001.

[45] D. Yohannes, S. Sarwana, S.K. Tolpygo, A. Sahu, and V. Semenov, "Characterization of HYPRES' 4.5 kA/cm$^2$ & 8 kA/cm$^2$ Nb/AlOx/Nb fabrication processes," *IEEE Trans. Appl. Supercond*. vol. 15, pp. 90, 2005.

[46] The SQUID Handbook, Vol. I *Fundamentals and Technology of SQUIDs and SQUID Systems,* (eds. John Clarke, Alex I. Braginski) Wiley-VCH, Weinheim, Germany, 2004.

[47] The SQUID Handbook, Vol. II *Applications of SQUIDs and SQUID Systems,* (eds. John Clarke, Alex I. Braginski) Wiley-VCH, Weinheim, Germany, 2006.

[48] S. Berggren, G. Prokopenko, P. Longhini, A. Palacios, et al, "Development of 2D bi-SQUID arrays with high linearity," *IEEE Trans. Appl. Supercond.*, vol. 23, 2013 (submitted).